\documentclass[usenatbib,useAMS]{mn2e}
\usepackage{natbib}
\usepackage{amsmath}
\usepackage{graphicx}
\usepackage{color}
\usepackage{epsfig}
\allowdisplaybreaks[1]

\newcommand{\msun}{{\rm M}_\odot}

\newcommand{\zsun}{Z_\odot}
\newcommand{\cc}{{\rm cm}^{-3}}

\newcommand{\msunyr}{{\rm M}_\odot~{\rm yr}^{-1}}

\newcommand{\K}{{\rm K}}
\newcommand{\beq}{\begin{equation}}
\newcommand{\eeq}{\end{equation}}

\title[SMS formation in HV collisions of protogalaxies]
{Direct collapse black hole formation via high-velocity collisions of protogalaxies}

\author[]{Kohei Inayoshi$^{1}$\thanks{E-mail: inayoshi@astro.columbia.edu}
\thanks{Japan Society for the Promotion of Science Postdoctoral Fellow},
Eli Visbal$^{1}$\thanks{E-mail: visbal@astro.columbia.edu}\thanks{Columbia Prize Postdoctoral Fellow in the Natural Sciences},
Kazumi Kashiyama$^{2}$\thanks{E-mail: kashiyama@berkeley.edu}\thanks{Einstein Fellow}\\
$^{1}$Department of Astronomy, Columbia University, 550 W. 120th Street, New York, NY 10027, USA\\
$^{2}$Department of Astronomy; Theoretical Astrophysics Center; University of California, Berkeley, Berkeley, CA 94720, USA}

\begin{document}

\maketitle

\label{firstpage}

\begin{abstract}
We propose high-velocity collisions of protogalaxies as a new pathway to form supermassive stars~(SMSs) 
with masses of $\sim 10^5 \ M_\odot$ at high redshift ($z > 10$).
When protogalaxies hosted by dark matter halos with a virial temperature of $\sim 10^4 \ \rm K$ collide 
with a relative velocity $\ga 200$ km s$^{-1}$, 
the gas is shock-heated to $\sim 10^6$ K and subsequently cools isobarically via 
free-free emission and He$^+$, He, and H line emission. 
Since the gas density ($\ga 10^4~\cc$) is high enough to destroy H$_2$ molecules by collisional dissociation, 
the shocked gas never cools below $\sim 10^4$ K. Once a gas cloud of $\sim 10^{5} \ \rm M_\odot$ reaches this temperature, 
it becomes gravitationally unstable and forms a SMS which will rapidly collapse into a super massive black hole (SMBH) 
via general relativistic instability. 
We perform a simple analytic estimate of the number density of direct-collapse black holes (DCBHs) formed 
through this scenario (calibrated with cosmological N-body simulations) and find $n_{\rm DCBH} \sim 10^{-9} \ \rm Mpc^{-3}$ 
(comoving) by $z = 10$. This could potentially explain the abundance of bright high-$z$ quasars. 
\end{abstract}



\section{Introduction}
The existence of bright quasars~(QSOs) at $z \ga 6-7$ 
\citep{2006NewAR..50..665F,2010AJ....139..906W,2011Natur.474..616M,2015Natur.518..512W}
presents an intriguing question: how do supermassive black holes (SMBHs) 
with masses $\ga {\rm a ~few} \times 10^9 ~M_\odot$ form within the first billion years after the Big Bang?
Perhaps the simplest possible explanation is that the $10-100 \ \msun$ black hole remnants of the first generation of stars grow into these supermassive black holes via gas accretion. However, this requires essentially uninterrupted Eddington limited accretion for the entire history of the Universe, which seems unlikely due to radiative feedback from the accreting BH 
\citep[e.g.][]{2007MNRAS.374.1557J,2009ApJ...696L.146M,2011ApJ...739....2P,2012ApJ...747....9P,2012MNRAS.425.2974T}.
Major mergers of BHs are not expected to accelerate growth significantly. 
This is because the kick velocity of the merged BH is typically larger than the escape velocity of its host galaxy
\citep[e.g.][]{2007ApJ...661..430H,2007PhRvL..99d1102K} leading to ejection and halting gas accretion
(but see also \citealt{TH09}).

An alternative scenario is the formation of supermassive stars~(SMSs) with a mass of $\ga 10^{5} \ \msun$
\citep{1994ApJ...432...52L,2003ApJ...596...34B,2004MNRAS.354..292K,2006MNRAS.370..289B},  
which directly collapse into SMBHs via general relativistic instability
\citep{1964ApJ...140..417C,1971reas.book.....Z,2002ApJ...572L..39S}. The larger mass of these seed BHs reduces the accretion time necessary to reach the masses implied by high-redshift QSOs.
Formation of a SMS requires a  $\sim 10^{5} \ \msun$ metal-poor gas cloud with no molecular hydrogen (H$_2$)
in a massive dark matter halo with virial temperature of $\sim 10^4$ K \citep{2003ApJ...596...34B}.
In the absence of H$_2$, the gas cloud can only cool by atomic hydrogen 
(Ly$\alpha$, two-photon, and H$^-$ free-bound emissions)
and the temperature remains $\sim 8000$ K \citep[e.g.,][]{O01}. 
Once such a massive gas cloud is assembled, it collapses monolithically and 
isothermally without significant fragmentation 
\citep{2003ApJ...596...34B, 2010MNRAS.402.1249S, 2009MNRAS.393..858R, 2009MNRAS.396..343R,
2013MNRAS.433.1607L, 2014MNRAS.445L.109I,2015MNRAS.446.2380B}.
After the collapse, a single protostar with a mass of $\sim 1~\msun$ is formed at the center of the cloud. 
The protostar grows via rapid gas accretion, $\ga 1~\msunyr$~\citep{2014MNRAS.445L.109I,2015MNRAS.446.2380B}, 
and becomes a SMS within $\sim 1$ Myr~\citep{2012ApJ...756...93H,2013ApJ...778..178H,2013A&A...558A..59S}.

Suppressing molecular hydrogen cooling is the largest obstacle to high-redshift SMS formation. One way to achieve this is through H$_2$ photodissociation by far-ultraviolet (FUV) photons in the Lyman-Werner (LW) band ($11.2-13.6$ eV)
\citep[e.g.,][]{O01,2002ApJ...569..558O, 2003ApJ...596...34B,2010MNRAS.402.1249S,IO11,
2014MNRAS.443.1979L,2014MNRAS.445..544S}.
In order to dissociate H$_2$ in a massive dark matter halo, the required FUV intensity is $J_{\rm LW}^{\rm crit}\simeq 1500$
\citep[in units of $10^{-21}$ erg s$^{-1}$ cm$^{-2}$ sr$^{-1}$ Hz$^{-1}$;][]{2014MNRAS.445..544S}. 
Given that star-forming galaxies are also strong X-ray sources, $J_{\rm LW}^{\rm crit}$ may increase by a factor $\sim 10$
\citep{IO11,2014arXiv1411.2590I}. 
This is due to the electron-catalyzed reactions (${\rm H}+{\rm e}^- \rightarrow  {\rm H}^-+\gamma$; 
$\rm{H}^- + {\rm H}\rightarrow \rm{H}_2+\rm{e}^-$),
which promote H$_2$ formation in low-metallicity gas. 
Massive dark matter halos irradiated with $J_{\rm LW}^{\rm crit} \sim 10^4$ are expected to be extremely rare. 
Direct-collapse BHs~(DCBHs) formed in this way may not be able to account for the abundance of observed high-$z$ quasars 
\citep{Dijkstra+14,2014arXiv1411.2590I}. 
We point out that a variation of this scenario based on the synchronized formation and merger of pairs of 
$T_{\rm vir}\sim 10^4 ~ \rm{K}$ dark matter halos, may produce enough DCBHs to explain the observations 
\citep{2014MNRAS.445.1056V}. 
However additional numerical simulations are required to validate this possibility.

A different pathway to H$_2$ suppression is collisional dissociation (H$_2$ + H $\rightarrow$ 3H), 
which can occur if the metal poor gas reaches high density and temperature, satisfying  
$(n/10^2 \ {\rm cm^{-3}}) \times (T/10^6 \ {\rm K}) \ga 1$ (the so-called ``zone of no return'').  
\cite{IO12} proposed that galactic-scale shocks can satisfy this condition.
If the shock happens at the central region of a massive halo $\la 0.1~R_{\rm vir}$, 
the density and temperature of the shocked gas become $n\ga 10^4~\cc$ and $T\ga 10^4$ K, 
and efficient collisional dissociation of H$_2$ can occur. 
However, the simulations of \cite{2014MNRAS.439.3798F} showed that for several examples 
of less massive halos with $T_{\rm vir}\la 10^4$ K, shocks do not reach the center preventing SMS formation. 
It was also pointed out that in typical halos (not the high-velocity collisions discussed in this paper), the zone of no return cannot 
be reached without radiative cooling, which may lead to star formation and prevent SMS formation \citep{Visbal+14}. 
SMS formation may still be possible in larger halos ($T_{\rm vir}\ga 10^4$ K) if shocks reach the centers of the halos
before significant amounts of stars of formed. This requires further study with numerical simulations.

Another proposed SMS formation scenario is based on massive-galaxy mergers~\citep{2010Natur.466.1082M, Mayer_et_al_2014}. 
A merger can drive strong gas inflow and supersonic turbulence in the inner galactic core, 
which prevents significant fragmentation of the gas even with some metals~\citep[but see][]{2013MNRAS.434.2600F}. 
If efficient angular momentum transport can be sustained in the inner $\sim 0.1 \ \rm pc$ for a sufficiently long time (which requires confirmation from further numerical simulations) a SMS of $\ga 10^{8} \ \msun$ could form. 
DCBHs from such SMSs might explain the observed abundance of high-$z$ QSOs.

In this paper, we propose high-velocity collisions of protogalaxies as a new pathway to form SMSs and DCBHs at high redshift. 
As observed in the local Universe, a fraction of galaxies and also clusters of galaxies collide with a much larger velocity than the
typical peculiar velocity (e.g., the Taffy galaxy and the bullet cluster; 
\citealt{1993AJ....105.1730C,2002AJ....123.1881C,1998ApJ...496L...5T,2002ApJ...567L..27M}). 
At the interface of such colliding galaxies, shock-induced starbursts have been confirmed
\citep[e.g.,][]{1978ApJ...219...46L, Saitoh_et_al_2009}. 
We show that, when a similar collision with a high-velocity $\ga 200$ km s$^{-1}$ happens between metal-poor galaxies, 
a hot gas ($\sim 10^6$ K) forms in the post-shock region and the subsequent radiative cooling makes the gas dense enough
that any H$_2$ molecules are destroyed by collisional dissociation.
Once a gas clump of $\sim 10^5~\msun$ with a low concentration of H$_2$ due to collisional dissociation forms,
its gravitational collapse can be triggered and a SMS forms.
Note that our scenario does not require supersonic turbulence or extremely efficient angular momentum transfer 
as in the galaxy merger scenario. We estimate the abundance of SMSs and DCBHs produced by high-velocity galaxy collisions, 
and show that it can be comparable to that of high-$z$ QSOs.

This paper is organized as follows. 
In \S2, we derive the necessary conditions to form SMSs in protogalaxy collisions.
In \S3, we estimate the number density of protogalaxy collisions resulting in the SMS formation,  
and show that the DCBHs from such SMSs could be the seeds of high-redshift QSOs.
Finally, we summarize and discuss our results in \S4.
Throughout we assume a $\Lambda$CDM cosmology consistent with the latest constraints from \emph{Planck} \citep{2014A&A...571A..16P}: $\Omega_\Lambda=0.68$, $\Omega_{\rm m}=0.32$, $\Omega_{\rm b}=0.049$, $h=0.67$, $\sigma_8=0.83$, and $n_{\rm s} = 0.96$.


\section{SMS formation via protogalaxy collisions } 
Generally speaking, a SMS can form from a $\ga 10^{5} \ \msun$ metal-poor gas clump without H$_2$.
Since H$_2$ can form via the electron-catalyzed reactions 
(${\rm H}+{\rm e}^- \rightarrow {\rm H}^- + \gamma; {\rm H}^- + {\rm H} \rightarrow {\rm H}_2 + {\rm e}^-$), 
it must be efficiently dissociated.  
We propose a SMS formation scenario where H$_2$ is dissociated from the shocks produced by a high-velocity collision of two dark matter halos. In this section, we show that SMS formation requires the relative velocity of the colliding protogalaxies to be in a specific range. If the collision velocity is too low, the gas will not be shocked to sufficient temperature and density to dissociate H$_2$. On the other hand, if the velocity is too high and the shock too violent, the gas will be disrupted before it can cool via atomic hydrogen and form a SMS. This velocity window depends on redshift due to the evolution of the typical gas properties of pre-shocked gas within dark matter halos.

\subsection{Protogalaxy properties}
Next, we describe the properties of dark matter halos and the gas within them as a function of redshift. 
This sets the collision velocity bounds for SMS formation which are derived below.
We consider protogalaxies hosted by dark-matter halos with virial temperatures of $T_{\rm vir} \sim 10^4$ K, corresponding to the atomic cooling threshold. 
Larger halos undergo radiative cooling which triggers star formation and metal enrichment, inhibiting SMS star formation ~(see Sec. \ref{sec:metal_enrichment} for further discussion).  
On the other hand, for halos $T_{\rm vir} \ll 10^4 \ \rm K$, the small gas mass is not sufficient to form a SMS~(see equation \ref{eq:Mgas}).

The virial mass of a dark matter halo is given by  
\begin{equation}
M_{\rm cool} \simeq 1.7 \times 10^7~\msun \left(\frac{T_{\rm vir}}{10^4~{\rm K}}\right)^{3/2}
\left(\frac{1+z}{16}\right)^{-3/2},
\label{eq:Mcool}
\end{equation}
and the virial radius by  
\begin{equation}
R_{\rm vir}\simeq 350~h^{-1}~{\rm pc}~\left(\frac{T_{\rm vir}}{10^4~\K}\right)^{1/2}
\left(\frac{1+z}{16}\right)^{-3/2}
\label{eq:Rvir}
\end{equation}
\citep{2001PhR...349..125B}. 

Simulations show that before cooling becomes efficient the central regions of dark matter halos contain a gas core with 
approximately constant density and radius \citep[see e.g.][]{Visbal+14}
\begin{equation}
R_{\rm core} \simeq 0.1~R_{\rm vir}.
\label{eq:Rcore}
\end{equation}
The gas core is surrounded by an envelope with a density profile roughly given by $\propto r^{-2}$.
The entropy profile, defined as $K=k_{\rm B}Tn_0^{-2/3}$, also has a core with $K/K_{\rm vir} \sim 0.1$. 
Here $K_{\rm vir}=k_{\rm B}T_{\rm vir}\bar n_{\rm b}^{-2/3}$ and 
$\bar n_{\rm b}$ is $200\Omega_{\rm m}^{-1}$ times mean number density of baryons
\citep[e.g.,][]{Visbal+14}.
Since $T\simeq T_{\rm vir}$ in the (pre-shock) gas core, the gas density can be estimated as 
\begin{equation}
n_0 \simeq \left(\frac{K}{K_{\rm vir}}\right)^{-1.5}{\bar n_b}
\simeq 22~\cc\left(\frac{1+z}{16}\right)^3.
\label{eq:pre_n}
\end{equation}
The total core gas mass is 
\begin{equation}\label{eq:Mgas}
M_{\rm gas, core} \sim 3.0\times 10^5~\msun~\left(\frac{T_{\rm vir}}{10^4~{\rm K}}\right)^{3/2}\left(\frac{1+z}{16}\right)^{-3/2},
\end{equation}
which is $\sim 10$ per cent of the total gas mass inside the dark-matter halo.

\subsection{Lower velocity limit}
The lower collision velocity limit is set by the requirement that H$_2$ is collisionally dissociated.
This will happen if the shock is strong enough for the gas to enter the so called  ``zone-of-no-return"~\citep[see][and Appendix A]{IO12}.
This region in temperature-density space is defined by  
\begin{equation}
T\ga 5.2\times 10^5~\K \left(\frac{n}{100~\cc}\right)^{-1},
\label{eq:zone}
\end{equation}
for $n\la10^4~\cc$, where $T$ and $n$ are the post-shock temperature and density, respectively.
For a collision velocity, $v_0$, much larger than the sound speed of the pre-shocked gas ($\sim 10$ km s$^{-1}$), 
\begin{equation}
T = \frac{3\mu m_{\rm p} v_0^2}{16k_{\rm B}} \simeq 8.5 \times 10^5~{\rm K}\left(\frac{v_0}{250~{\rm km~s^{-1}}}\right)^2,
\label{eq:post_T}
\end{equation}
and 
\begin{equation}
n = 4n_0,
\label{eq:post_n}
\end{equation}
where we set the mean molecular weight as $\mu=0.6$.
From equations (\ref{eq:pre_n}), (\ref{eq:zone}), (\ref{eq:post_T}), and (\ref{eq:post_n}), 
one can obtain the lower limit of the collision velocity for SMS formation, which is given by 
\begin{equation}
v_0 \ga 210 \ {\rm km \ s^{-1}} \ \left(\frac{1+z}{16}\right)^{-3/2}.
\label{eq:lower_limit}
\end{equation}

\begin{figure}
\begin{center}
\includegraphics[height=58mm,width=80mm]{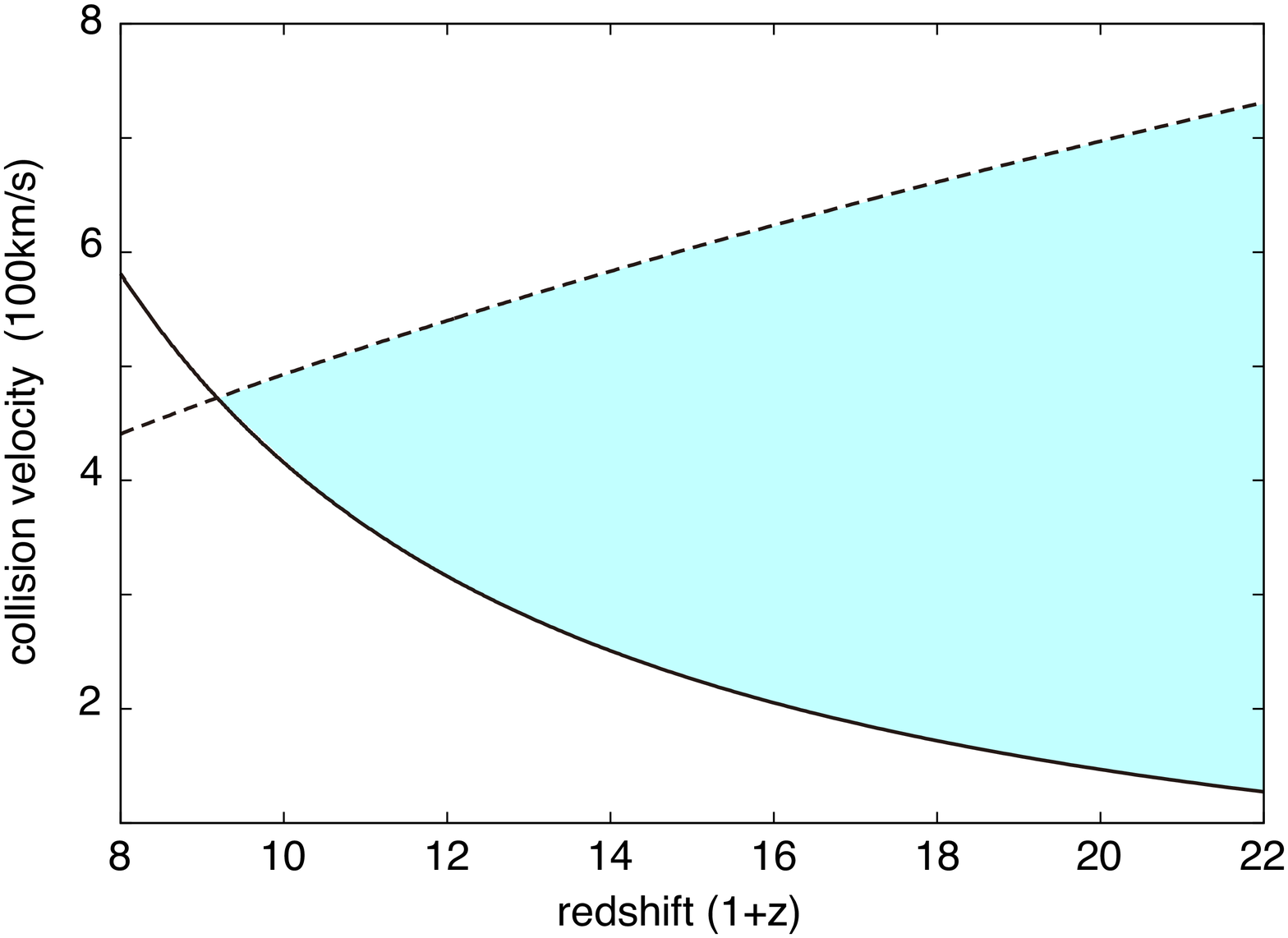}
\caption{
The collision-velocity window for supermassive star formation in protogalaxy collisions (shaded light-blue region). 
The solid curve shows the lowest velocity required to induce collisional dissociation of H$_2$~(equation \ref{eq:lower_limit}).
The dashed curve shows the highest velocity that meets the radiative shock condition~(equation \ref{eq:upper_limit}).
}
\label{fig:z_v}
\end{center}
\end{figure}

\subsection{Upper velocity limit}
The upper collision velocity limit leading to SMS formation is set by the radiative shock condition.
If the collision velocity is too large, the shocked gas starts to expand adiabatically before the radiative cooling sets in. 
In this case, the shocked gas cannot become dense enough to dissociate H$_2$ collisionally, and the shock-induced SMS formation cannot be triggered.
Therefore, the radiative cooling time needs to be shorter than the dynamical time of the shock. 

The dynamical time of the shock can be estimated as 
\begin{equation}
t_{\rm dyn} = \frac{R_{\rm core}}{v_0} \simeq 2.0 \times 10^5~{\rm yr}
\left(\frac{R_{\rm core}}{50~{\rm pc}}\right) \left(\frac{v_0}{250~{\rm km~s^{-1}}}\right)^{-1}, 
\label{eq:t_sc}
\end{equation}
while the radiative cooling time of the shocked gas is given by
\begin{equation}
t_{\rm cool}=\frac{3nk_{\rm B}T}{2\Lambda_{\rm rad}}. 
\label{eq:t_cool_gene}
\end{equation}
Here, $\Lambda_{\rm rad}(n,T)=n^2\bar \Lambda (T)$ is the cooling rate (in units of erg s$^{-1}$ cm$^{-3}$). 
The cooling function $\bar \Lambda(T)$ consists of the contributions from 
atomic H line emission at $T\sim 10^4$ K, atomic He$^+$ and He line emissions at $T\sim 10^5$ K, 
and the bremsstrahlung emission $\propto T^{1/2}$ at $T\ga 10^6$ K 
\citep{1993ApJS...88..253S,2007ApJ...666....1G}. 
In our scenario, the shocked gas temperature initially ranges up to $5\times 10^5 {\ \rm K} \la T \la 5\times 10^6$ K, 
which corresponds to the collision velocity of $150 \ {\rm km \ s^{-1}} \la v_0 \la 500$ km s$^{-1}$ (see equation \ref{eq:post_T}). 
For such gas, the cooling function can be well approximated by $\bar \Lambda_0 \simeq  5\times 10^{-24}$ erg s$^{-1}$ cm$^3$. 
Given that $n \propto T^{-1}$ during the radiative contraction, the cooling time becomes shorter for a lower-temperature, 
i.e. the gas contracts in a thermally unstable way. 
Thus, the radiative cooling time is essentially given by substituting $\bar \Lambda_0$ into equation (\ref{eq:t_cool_gene}); 
\begin{equation}
t_{\rm cool,max}\simeq 2.8 \times 10^4~{\rm yr}~\left(\frac{v_0}{250~{\rm km~s^{-1}}}\right)^2 \left(\frac{n_0}{10~\cc}\right)^{-1}.
\label{eq:t_cool}
\end{equation}
From equations (\ref{eq:Rvir} $-$ \ref{eq:pre_n}), (\ref{eq:t_sc}), and (\ref{eq:t_cool}), the radiative shock condition ($t_{\rm dyn}\ga t_{\rm cool}$) can be rewritten as 
\begin{equation}
v_0 \la 620 \ {\rm km \ s^{-1}} \ \left(\frac{T_{\rm vir}}{10^4~\K}\right)^{1/6} \left(\frac{1+z}{16}\right)^{1/2}.
\label{eq:upper_limit}
\end{equation}

Fig.~\ref{fig:z_v} shows the collision velocity window for SMS formation as a function of redshift (the shaded region). 
For collisions of $T_{\rm vir}\sim 10^4$ K halos in this window, gas is shocked into the zone-of-no-return and H$_2$ molecules are destroyed by collisional dissociation. Additionally, the shocked gas radiatively cools to $\sim 10^{4} \ \rm K$ within the shock dynamical time. 
Once the $\ga 10^{5} \ \msun$ gas cloud is assembled and cools, it becomes unstable 
due to self-gravity, and a SMS can be formed~\citep[e.g.][]{2014MNRAS.445L.109I,2015MNRAS.446.2380B}. 
Note that since the total mass of the colliding gas in dark matter halos with $T_{\rm vir}\simeq 10^4$ K
can be as large as $\sim 10^{6} \ \rm \msun$, 
several SMSs may be formed at once in the collisions we consider. 
These SMSs would result in DCBHs of $\ga 10^{5} \ \msun$ at $z > 10$ , and as we show in the following section could potentially be the seeds of high-$z$ QSOs.


\section{DCBH Abundance from High-Velocity Collisions}
Precisely estimating the number of collisions which result in SMSs and DCBHs is very challenging because 
it depends on detailed nonlinear physics. This most likely necessitates N-body simulations, 
however, the rarity of these collisions (we estimate $\sim 10^{-9}~{\rm Mpc^{-3}}$ from $z=10-20$) requires simulations much larger than are feasible with current computers. 
Here we address this issue by performing a simple order of magnitude estimate with an analytic formula based on idealized assumptions and calibrated with a N-body simulation. 
We find that the number density of DCBHs could be high enough to explain observations of high-$z$ QSOs. 
However, we emphasize that our estimate has large uncertainties which we discuss in \S4.

\subsection{Collision rate}
We estimate the high-velocity protogalaxy collision rate by considering 
the number of dark matter halos just below the atomic cooling 
threshold that collide with a relative velocity in the range shown in Fig.~\ref{fig:z_v}.
For simplicity, we consider one halo moving with a very high peculiar velocity and the other with a typical velocity (determined with the N-body simulation described below) in the opposite direction.

Making the idealized assumption that halo positions and velocities are randomly distributed (i.e. ignoring clustering and coherent velocities, which are discussed in Sec. \ref{sec:rate}), the collision rate per volume is given by
\begin{equation}
\frac{dn_{\rm coll}}{dt} = n_{\rm fast} n_{\rm h} \times v_{\rm fast} \times \pi b^2
= f_{\rm fast} n_{\rm h}^2 \times v_{\rm fast}  \times \pi b^2,
\end{equation}
where $n_{\rm h}$ is the number density of all halos near the cooling threshold, 
$v_{\rm fast}$ is the velocity of the fast-moving halo necessary to form one (or several) SMS(s), 
$n_{\rm fast}=f_{\rm fast}n_{\rm h}$ is the number density of halos with peculiar velocity greater 
than this value (but below the maximum value), and $b$ is the impact parameter required for SMS formation. 
Note that these values are all initially calculated in physical units and the abundance is later converted to comoving units to compare with QSO observations.
We compute the halo number density with the Sheth-Tormen mass function \citep{1999MNRAS.308..119S} and consider a mass range of $(0.5-1)\times M_{\rm cool}$~(equation \ref{eq:Mcool}). 
For the impact parameter, $b$, we use ten per cent of the virial radius (equation \ref{eq:Rcore}).  
We assume $v_{\rm fast}=v_{\rm min} - v_{\rm typ}$, where $v_{\rm typ}$ is the typical halo peculiar velocity (assumed to be $40~{\rm km~s^{-1}}$) 
and $v_{\rm min}$ is shown in Fig.~\ref{fig:z_v} (solid curve). The fraction of fast-moving halos is given by
\begin{equation}
f_{\rm fast} = \int_{v_{\rm min}-v_{\rm typ} }^{v_{\rm max}-v_{\rm typ} } p(v) dv,
\end{equation}
where $v_{\rm max}$ is given in Fig.~\ref{fig:z_v} (dashed curve) and $p(v)$ is the peculiar velocity probability 
density function (PDF) which we estimate from a N-body simulation as described in the following subsection.

\begin{figure}
\begin{center}
\includegraphics[height=60mm,width=80mm]{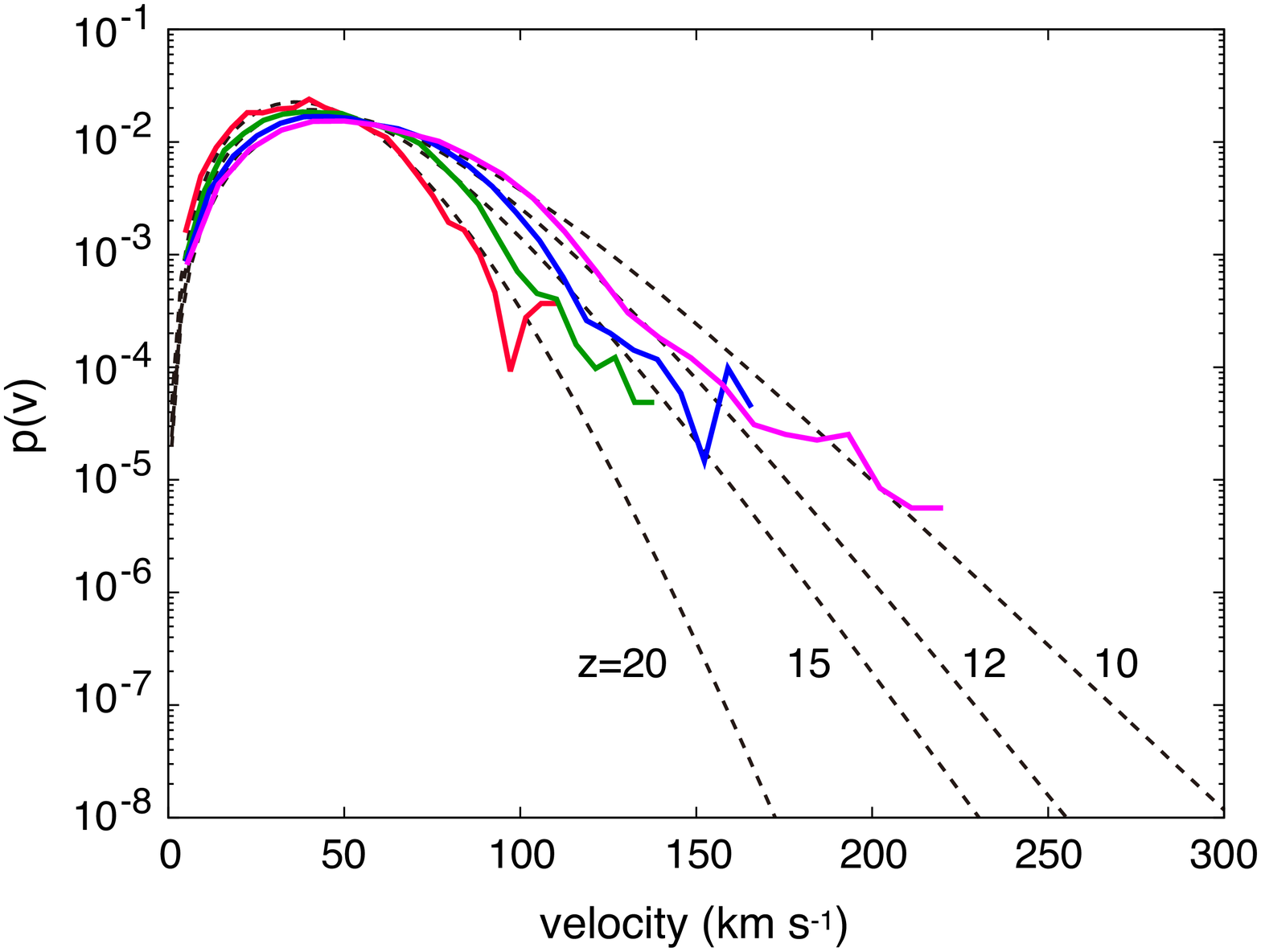}
\caption{The peculiar velocity PDF, $p(v)$, measured from our N-body simulation (solid curves) 
and the best fits described in Section 3 (dashed curves) for redshifts $z=20$, 15, 12, 10 (from left to right).}
\label{v_pdf}
\end{center}
\end{figure}

\subsection{N-body simulation and velocity PDF}
To estimate the dark matter halo peculiar velocity PDF, we ran a cosmological N-body simulation 
with the publicly available code \textsc{gadget2} \citep{2005MNRAS.364.1105S}. 
The simulation has a box length of length $10~h^{-1}$ Mpc (comoving) and mass resolution of $768^3$ particles, 
corresponding to a particle mass of $1.96\times10^5~h^{-1}~\msun$. Snapshots were saved at $z=20$, 15, 12, and 10.
We used the \textsc{rockstar} halo finder \citep{2013ApJ...762..109B} to locate dark matter halos and determine their masses and velocities.

\begin{figure*}
\begin{center}
\includegraphics[height=60mm,width=80mm]{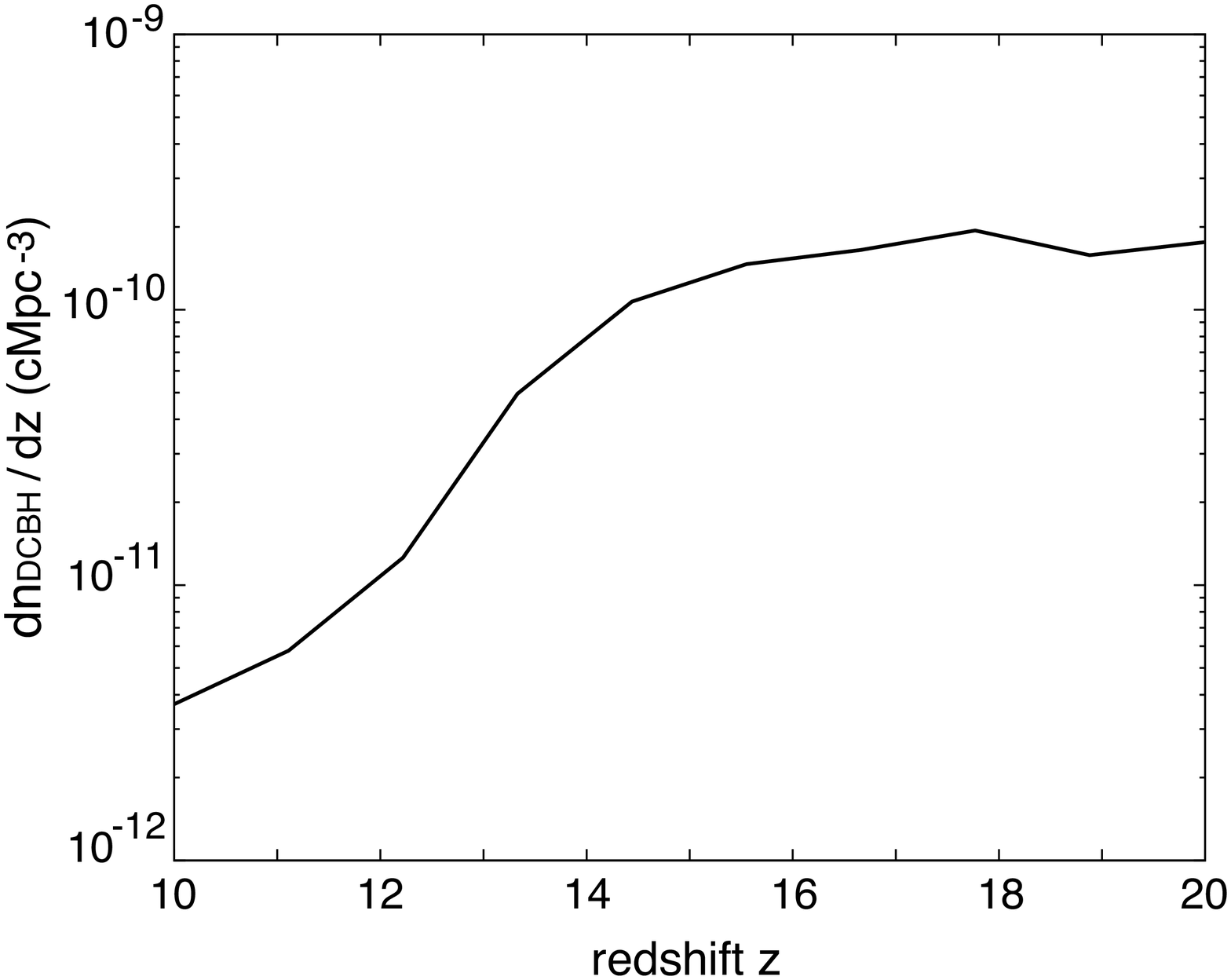}
\hspace{4mm}
\includegraphics[height=60mm,width=80mm]{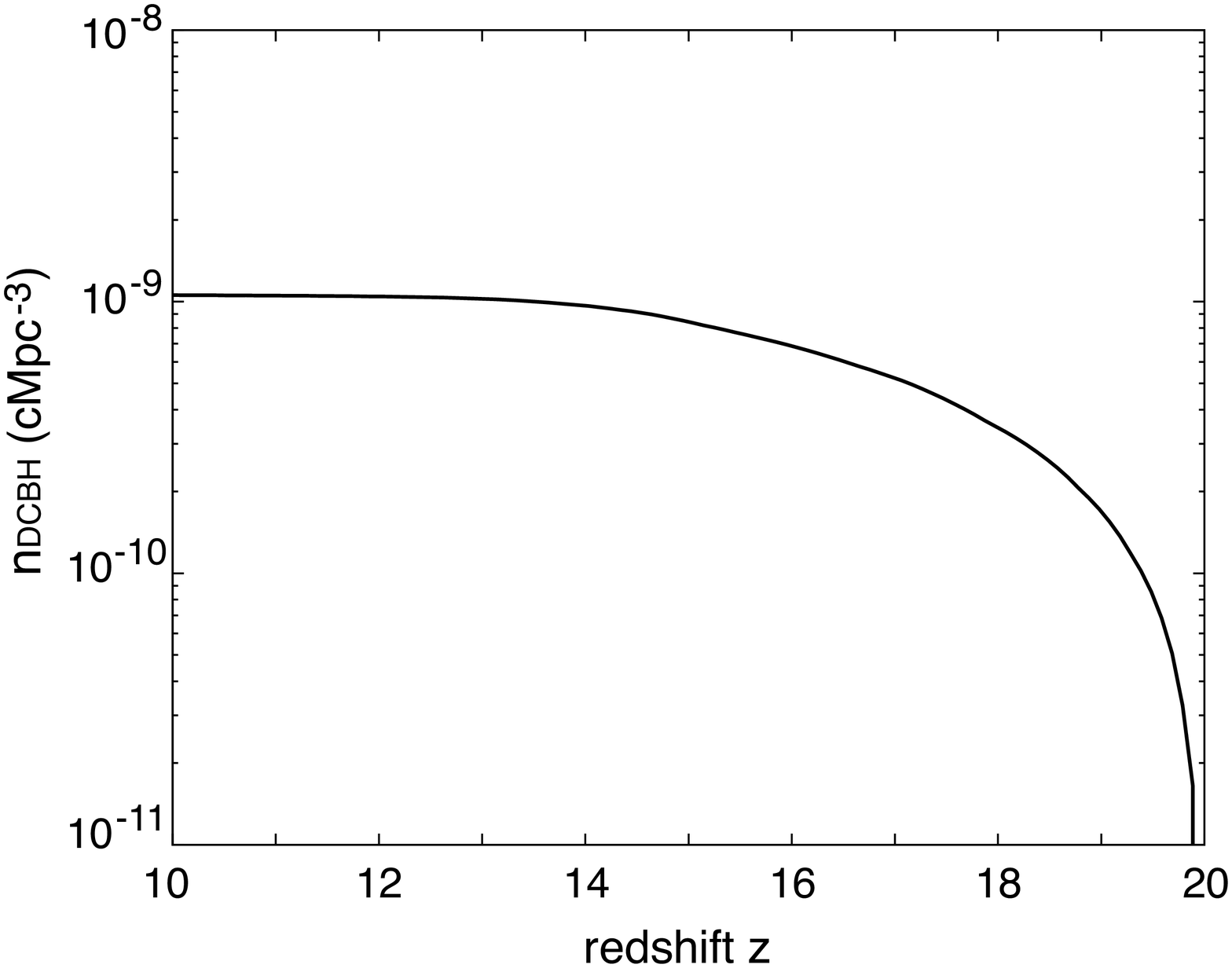}
\caption{The estimated number density of DCBHs created through high-velocity collisions per unit redshift (left panel) and the cumulative number density as a function of redshift (right panel).}
\label{dc_abund}
\end{center}
\end{figure*}

In Fig.~\ref{v_pdf}, we plot the velocity PDF of $M=(0.5-1)\times M_{\rm cool}$ dark matter halos at $z=20$, 15, 12, and 10.
We also plot fits with a form guided by \cite{2001MNRAS.322..901S} and \cite{2003MNRAS.343.1312H} 
who argue that the PDF is a Gaussian distribution in each velocity component 
(Maxwell-Boltzmann distribution for the total 1D velocity) at fixed halo mass and local overdensity. 
For halos near the cooling threshold this leads to
\begin{equation}
p(v) = \frac{\int d\delta p(\delta) (1+b_{\rm h} \delta) p(v | \delta)}{\int d\delta p(\delta) (1+b_{\rm h} \delta)},
\end{equation}
where $p(\delta)$ is the cosmological overdensity PDF, $b_{\rm h}$ is the Sheth-Tormen dark matter halo bias \citep{1999MNRAS.308..119S}, and $ p(v | \delta)$ is the velocity PDF at fixed $\delta$. 
The overdensity PDF is assumed to be a lognormal distribution \citep[see e.g.][]{1991MNRAS.248....1C}
\begin{equation}
p(\delta) = \frac{1}{\sqrt{2\pi \sigma_\delta^2}} \exp{\left (- \frac{[\ln(1+\delta) +  \sigma_\delta^2/2 ]^2}{2 \sigma_\delta^2} \right )} \frac{1}{1+\delta}.
\end{equation}
The velocity PDF at fixed overdensity is assumed to be a Maxwell-Boltzmann distribution with variance
\begin{equation}
\sigma^2 = \left ( 1  + \delta \right )^{2 \mu} \sigma_{v}^2.
\end{equation}
 We set $\sigma_\delta^2 = \ln[1 + 0.25/(1+z)]$ \citep{2003MNRAS.343.1312H} (which determines the size of the region corresponding to $\delta$) and fit two parameters to our data, $\sigma_{v}$ and $\mu$.
For redshifts of $z=20$, 15, 12, and 10, we find $\sigma_{v}=$24.04, 27.27, 30.91, and 33.54 and $\mu=$ 0.8687, 1.2404, 1.0949, and 1.2081, respectively.

\subsection{DCBH number density}
Using the best fit $p(v)$ discussed above, we find the number of DCBHs produced from $z=10-20$.
We assume that the LW background at $z<20$ suppresses 
star formation and subsequent metal enrichment in halos below the atomic cooling threshold.
At higher redshift we assume that star formation in minihalo progenitors prevents DCBH formation. 
The total number density formed as a function of redshift is given by
\begin{equation}
n_{\rm DCBH}(z) =  \int dz \frac{dt}{dz} \frac{dn_{\rm coll}}{dt}.
\end{equation}
To get the velocity PDF at intermediate redshifts between our simulation snapshots, 
we linearly interpolate the results from Fig.~\ref{v_pdf}.
In Fig.~\ref{dc_abund}, we plot the number density of DCBHs as a function of redshift. 
Most DCBHs come from high redshift due to the lower minimum velocity given in Fig.~\ref{fig:z_v}.
We find a total density by $z=10$ of $10^{-9} ~\rm{Mpc}^{-3}$ (comoving). 
Thus, it seems possible that these DCBHs could potentially explain the abundance of high-$z$ QSOs.


\section{Discussion and Conclusions}
We have shown that high-velocity collisions of metal-poor galaxies may result in the formation of supermassive stars (SMSs). 
When dark matter halos with a virial temperature $\sim 10^4$ K collide with a relative velocity 
$\ga 200$ km s$^{-1}$,  gas is heated to very high temperature ($\sim 10^6$ K) in the shocked region. 
The shocked gas cools isobarically via free-free emission and forms a dense sheet ($\ga 10^4~\cc$).
In this dense gas, H$_2$ molecules are collisionally dissociated, and
the gas never cools below $\sim 10^4$ K. 
Such a clump of gas with mass $\sim 10^5~\msun$, once assembled, becomes gravitationally unstable and forms a SMS,
which would directly collapse into a black hole (DCBH) via general relativistic instability.
We estimated the abundance of DCBHs produced by this scenario
with a simple analytical argument calibrated with cosmological N-body simulations and found a
number density of $\sim 10^{-9}$ Mpc$^{-3}$ (comoving) by $z=10$. This is large enough to explain the abundance of high-redshift bright QSOs.

\subsection{Observational Signatures}
Next, we briefly discuss the possible observational signatures of SMSs formed through high-velocity collisions of protogalaxies. 
The temperature of the shocked gas in the collisions we consider above is $T \la 10^{6} \ \rm K$ (equation \ref{eq:post_T}). 
The gas cools initially via bremsstrahlung, then atomic He$^{+}$ and He line emissions, and finally atomic H line emission. 
The intrinsic bolometric luminosity can be estimated as 
$L_{\rm bol} \sim 10 M_{\rm gas, core} {v_0}^2/t_{\rm cool, max} \la 10^{43} \ \rm erg \ s^{-1}$
for our representative case (see equations \ref{eq:Mgas} and \ref{eq:t_cool}).  
Given that the colliding gas in dark matter halos is mostly neutral, 
the cooling radiation is reprocessed into various recombination lines, e.g., Ly$\alpha$, H$\alpha$, and He II $\lambda1640$.  
The H$\alpha$ and He II $\lambda1640$ emission lines are particularly interesting, since the intergalactic medium would be optically thin to them. 
If $\sim {\rm a \ few}$ per cent of the bolometric luminosity goes into these lines, 
which is reasonable~\citep[see, e.g.,][for numerical simulations of cooling radiation from hot metal-poor gas with $\sim 10^{5} \ \rm K$]{Johnson+11},  
the emission could be detected from $z \la 15$ by the Near-Infrared Spectrograh~(NIRSpec) on the James Webb Space Telescope (JWST) with an exposure time of $\ga 100 \ \rm h$. 
Due to the high cooling temperature, the ratio of He II $\lambda1640$ to H$\alpha$ flux is expected to be large,
which may make it distinguishable from other sources (e.g. population III galaxies). 
In principle it may also be possible to detect H Ly$\alpha$ emission from protogalaxy collisions. Ly$\alpha$ emission could constitute a large fraction of the bolometric luminosity (perhaps 10 per cent). 
However, even if a collision is observed in Ly$\alpha$ it may be difficult to distinguish from other objects such as accreting massive dark matter halos.

A detailed radiative transfer calculation is required to accurately predict the emission spectrum, 
which is beyond the scope of this paper. 
Even though protogalaxy collisions may be bright enough to observe, recent collisions are expected to be extremely rare. 
At most there will be $\sim$ a few in the whole sky, given the event rate, 
$dn_{\rm coll}/dt (z=15) \sim 10^{-11}$ Mpc$^{-3}$ (comoving) Myr$^{-1}$
and the emission duration, $\sim \ 0.1 \ \rm Myr$.
Thus, it will be extremely challenging to detect the signal described above in the near future.

\subsection{Impact of assumptions}
Here we discuss some of the key assumptions we made, and how changes to these assumptions would affect our results. 

\subsubsection{Metal enrichment}
\label{sec:metal_enrichment}

In \S2, we calculated the thermodynamics of the shocked gas after protogalaxy collisions assuming zero metallicity.
This assumption is valid for gas metallicity smaller than $\la 10^{-3}~\zsun$ \citep{IO12}. 
If the metallicity is higher than this critical value, the shocked gas can cool down to below $\sim 10^4$ K 
via metal-line emissions (C$_{\rm II}$ and O$_{\rm I}$) and fragment into clumps of $\sim 10~\msun$, preventing SMS formation. 

In general, the metal-enrichment of gas in massive dark matter halos proceeds in two different ways. 
The first is internal enrichment by in situ star formation. Although not yet completely understood, the earliest star formation is expected to be 
triggered by H$_2$ cooling in progenitor ``minihalos" with $T_{\rm vir}<10^{4} \ \rm K$, 
which eventually assemble into the more massive dark matter halos we consider in this paper.
The level of self enrichment in minihalos is sensitive to the initial mass function~(IMF) of population III stars
\citep[e.g.][]{2014ApJ...781...60H,2014ApJ...792...32S}. 
If the IMF is extremely top heavy, the metal enrichment is predominately provided by pair-instability SNe. 
In this case, the metallicity at the gas core inside the dark matter halo could be as large as $\sim 10^{-4}-10^{-3}~\zsun$ at 
$z \sim 10$ \citep[e.g.][]{2010ApJ...716..510G, 2012ApJ...745...50W}. 
On the other hand, if the IMF is mildly top heavy, core-collapse SNe from $\sim 40~\msun$ stars 
would be the dominant source \citep{2011Sci...334.1250H,2012MNRAS.422..290S,2012ApJ...760L..37H}.  
In this case, the metallicity may be one order of magnitude lower ($\la 10^{-4}~\zsun$) 
at the same redshift \citep{2002ApJ...567..532H, 2006NuPhA.777..424N}.

In our abundance estimates of SMS formation through high-velocity collisions, we only consider dark matter halos 
with $T_{\rm vir} \la 10^4 \ \rm K$. 
We assume that below $z=20$ the abundance of H$_2$ required for star formation in minihalos is sufficiently suppressed by LW background radiation ~\citep[e.g.][]{2000ApJ...534...11H,2001ApJ...548..509M,2007ApJ...671.1559W,2008ApJ...673...14O}. 
The required LW background flux for this to occur is estimated to be $J_{\rm LW} \sim 0.2-2~(3\times 10^{-4}-4\times 10^{-2})$ at $z=15~(20)$~\citep{2014MNRAS.445.1056V}. 
The anticipated LW background flux is $J_{\rm LW}\sim 0.1-10~(0.01-20)$ at $z\sim 15~(20)$
\citep[e.g.][]{2009ApJ...695.1430A,2013MNRAS.428.1857J,2014MNRAS.445..107V}, which depends on the detailed properties of Pop III stars and the efficiency with which they are produced. While there are certainly large uncertainties in the LW background, we find our assumption of minihalo star formation suppression to be reasonable.

The second way in which halos can obtain metals is through external enrichment by galactic winds from nearby massive galaxies. 
Semi-analytic models predict that the intergalactic medium can be polluted by this effect leading to an average metallicity of $\langle Z\rangle \simeq 10^{-4}~\zsun$ by $z\ga 12$ \citep[e.g.][]{2007MNRAS.382..945T,2010MNRAS.407.1003M}.
However,the fraction of the intergalactic medium  that has been polluted is expected to be small at the high redshifts important for our calculation (e.g the estimated volume filling factor is $\sim 10^{-4}$ for $z>12$; \citealt{2014MNRAS.440.2498P}). 
Thus, external metal enrichment is unlikely to impact our assumption of zero metallicity. 

In summary, it is reasonable to neglect the effect of metal cooling for dark-matter halos with a virial temperature $T_{\rm vir} \la 10^{4} \ \rm K$ at $z \ga 10$ after the LW background suppresses star formation in minihalos. 

\subsubsection{Gas thermodynamics}
In this paper, we derived the conditions for SMS formation in protogalaxy collisions (Fig. \ref{fig:z_v}) 
based on the``zone of no return" shown in Appendix A. 
This zone is obtained from a one-zone calculation of thermodynamics of the shocked gas. 
Of course, galaxy collisions are actually three-dimensional phenomena; detailed hydrodynamical simulations are necessary to confirm our scenario. 

We obtained equation (\ref{eq:zone}) by assuming that the shock is plane-parallel and steady. 
This assumption would be valid for nearly head-on collisions and timescales shorter
than the shock dynamical timescale. 
Accordingly, we set the maximum impact parameter as $b \sim 0.1~R_{\rm vir}$, 
which corresponds to the size of the gas core of an atomic-cooling halo.
However, galaxy collisions occur typically with a larger impact parameter $b \sim R_{\rm vir}$. 
A critical impact parameter for SMS formation needs to be identified by numerical simulation of protogalaxy collisions.
We note that the formation rate of SMSs and DCBHs in our scenario is somewhat sensitive to this critical value ($\propto b^2$). 

As mentioned in \S2, the shocked gas in the zone of no return is thermally unstable. 
Once the instability is triggered, fluctuations in the shocked gas grow and form clumpy structures 
with a length scale of $\la c_{\rm s}t_{\rm cool}$~\citep{1965ApJ...142..531F}. As a result of this, the structure of the shocked gas deviates from the plane-parallel sheet in a cooling time.
Unfortunately, our one-zone calculation cannot capture these effects.
Note that, as long as the H$_2$ abundance is suppressed, the cooling length is kept shorter than the Jeans length, 
thus the thermal instability does not necessarily result in a smaller fragmentation mass. 
Nevertheless, the effects of the thermal instability on SMS formation need to be studied using high-resolution simulations. 

We implicitly assumed that after radiative cooling of the shocked gas becomes irrelevant (i.e. $t_{\rm cool} \ga t_{\rm ff}$),
the corresponding Jeans mass is assembled, perhaps within $\sim t_{\rm ff}$, and the gas clump collapses due to its self-gravity. 
Our one-zone calculation cannot address how the mass assembly process actually proceeds in detail. 
Even when the mass budget is large enough $> 10^5~\msun$, the mass assembling may be halted, 
e.g., due to the angular momentum of the gas. A detailed numerical simulation is also required to clarify this point.

Subsequent mass accretion onto the DCBHs formed in protogalaxy collisions is also uncertain at this stage. 
This needs to be clarified in order to address whether such DCBHs can be the seeds of high-$z$ QSOs. 
The initial mass of the DCBHs is $\sim 10^5~\msun$ whereas the total gas mass of each colliding galaxy is $\ga 10^6~\msun$. 
We also note that the DCBH is unlikely to be hosted by the dark matter halo, at least just after the formation, 
because the collision velocity of the parent halos is much larger than the virial velocity.
Nevertheless, continuous mass accretion from the intergalactic medium may be expected since high-velocity collisions 
typically occur in over-dense regions.
Additional galactic and intergalactic-scale calculations including radiative feedback from accreting BHs are required to confirm this.

\subsubsection{High-velocity collision rate estimate}\label{sec:rate}
There are a number of uncertainties associated with the various assumptions we made 
to estimate the number density of DCBHs produced from high-velocity protogalaxy collisions. 
First, we note that our estimate depends strongly on the precise values of $v_{\rm min}$. 
Due to the steepness of $p(v)$ at large $v$, a 20 per cent decrease in $v_{\rm min}$ increases the abundance of DCBHs by more than an order of magnitude. 
The abundance also depends strongly on the value of the impact parameter needed to create a black hole ($n_{\rm DCBH} \propto b^2$). 
Future hydrodynamic simulations of individual collision events are needed to constrain $v_{\rm min}$ and $b$.

Additionally, the small box size of our simulation may systematically reduce the abundance of halos with high peculiar velocity. 
This is because large-scale density fluctuations, corresponding to modes larger than the box are artificially removed. 
We leave it to future work to determine how much this effect could enhance the number density we compute here.

\begin{figure}
\begin{center}
\includegraphics[height=60mm,width=80mm]{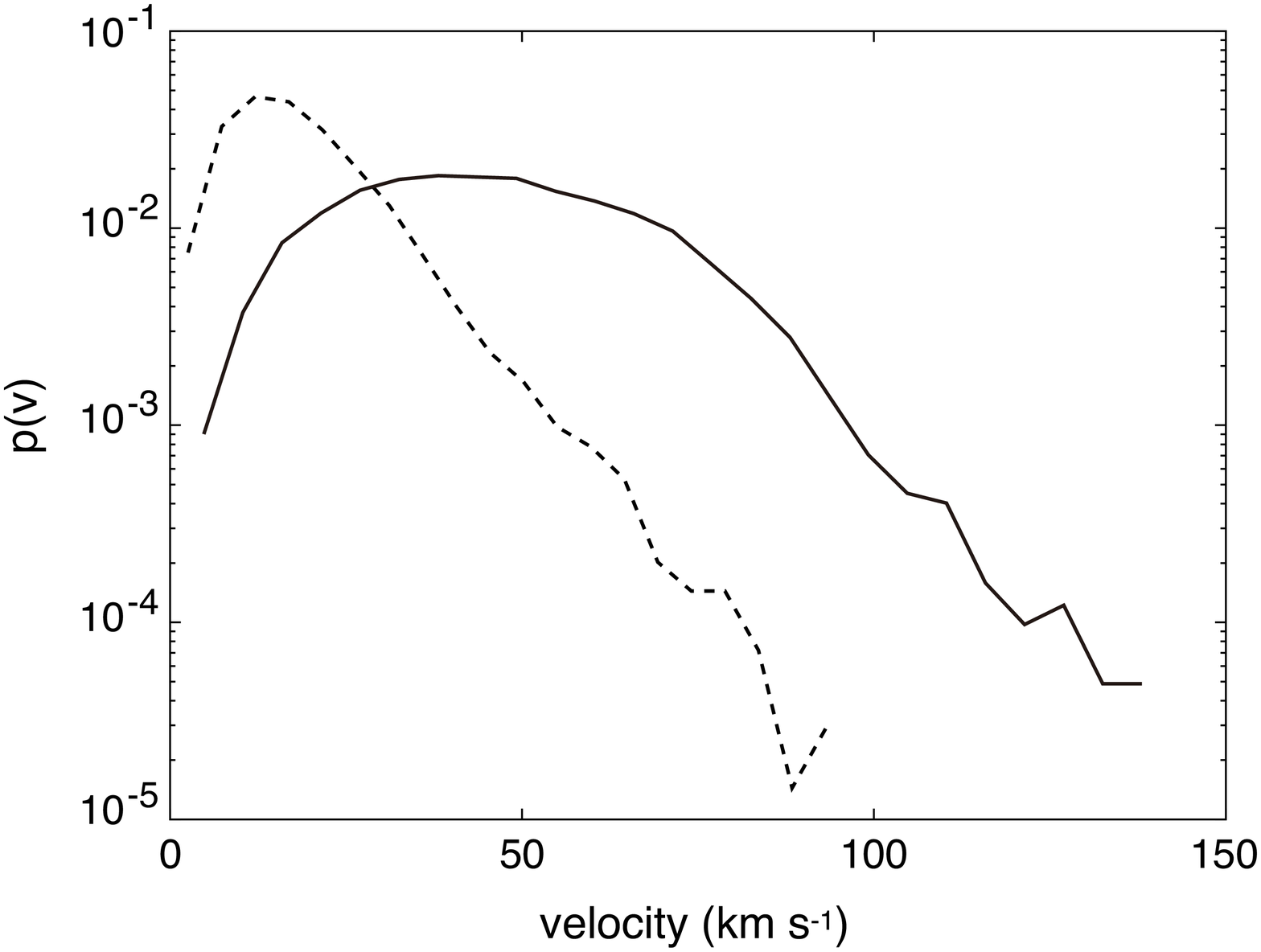}
\caption{The velocity PDF computed directly from our N-body simulation at $z=15$ (solid curve) and computed after subtracting away the local coherent velocity as described in Section 4.2.3 (dashed curve).}
\label{coh_pdf}
\end{center}
\end{figure}

We considered the case of one fast-moving halo and one halo at typical peculiar velocity in the opposite direction. 
Of course there can be other combinations of peculiar velocities and collision angles which lead to a DCBH. 
We find that we get similar results performing the more complicated analysis of adding up the contribution 
from all angles and different combinations of peculiar velocities. 
There is a factor of a few enhancement compared to the simple calculation discussed above. 

We note that our idealized assumptions of random positions and velocity directions are not expected to be accurate 
and estimate their impact on our number density estimate here. 
We expect that fast halos will preferentially be found in over dense regions, possibly regions which will soon virialize. 
The halo density enhancement (ignored in our estimate) is given by $(1+b_{\rm h} \delta)$ 
and the density of DCBHs will depend on this value squared. 
At $z=15$, in a region that is about to virialize ($\delta \approx \delta_{\rm c}=1.686$), 
the density enhancement of DCBHs is roughly 100. 

Our assumption of random velocities most likely overestimates the number density of DCBHs formed. 
This is because on small scales there will be some velocity coherence between nearby halos, reducing their relative velocities. 
To estimate the impact of this effect we recompute $p(v)$ from our simulation, 
and for each halo subtract away the mass-weighted mean velocity of all other nearby halos in the mass range $M=(0.5-1)M_{\rm cool}$. 
We include all halos within the typical separation length of these halos, $R_{\rm c}=n_{\rm h}^{-1/3}$. 
Increasing the value of $R_{\rm c}$ by a factor of two does not significantly affect our results. 
If this distance is taken to be significantly smaller there are not enough neighbors to compute the coherent velocity. 
This PDF at $z=15$ is shown in Fig.~\ref{coh_pdf}. 
At high $v$, it is more than an order of magnitude lower than $p(v)$ obtained without subtracting coherent velocities. 
The typical peculiar velocity is reduced by $\sim 25~{\rm km~s^{-1}}$. 
The relative changes at $z=20$ are similar. 
We find that these two effects (the high-$v$ $p(v)$ and the typical $v$) lower the abundance of DCBHs by approximately 
two orders of magnitude, which may roughly cancel when combined with the correction due to the density enhancement discussed above.

Despite the large uncertainties described above, the high-velocity collision of protogalaxies is an interesting pathway to form SMSs and DCBHs 
without extremely strong LW radiation and could explain the abundance of high-$z$ bright QSOs. 
In future work, we plan to perform detailed numerical studies on the gas dynamics of colliding galaxies and the event rate of appropriate collisions to determine if these events could truly be responsible for the first SMBHs.

\section*{Acknowledgements}
We thank Zolt\'an Haiman, Greg Bryan, Hidenobu Yajima, Eliot Quataert, P\'eter M\'esz\'aros and Renyue Cen for fruitful discussions.
This work is partially supported by the Grants-in-Aid by the Ministry of Education, Culture, and Science of Japan (KI), 
and by NASA through Einstein Postdoctoral Fellowship grant number PF4-150123 awarded by the Chandra X-ray Center, 
which is operated by the Smithsonian Astrophysical Observatory for NASA under contract NAS8-03060~(KK).  EV is supported by the Columbia Prize Postdoctoral Fellowship in the Natural Sciences.
Our N-body simulations were carried out at the Yeti High Performance
Computing Cluster at Columbia University.

\appendix

\section{zone of no return: Equation (5)}
Following \cite{IO12}, here we calculate thermal evolution of the shocked gas flow discussed above.
We assume that the flow is steady and plane-parallel \citep{1987ApJ...318...32S}, 
which is appropriate for almost head-on collisions in the shock dynamical time. 
The conservation of mass and momentum 
between the density $\rho_0$, pressure $p_0$ and flow velocity $v_0$ in the pre-shock flow
and those in the post-shock flow $\rho$, $p$ and $V$ give:
\begin{equation}
\rho V =\rho_0v_0,
\label{eq:app1}
\end{equation}
\begin{equation}
\rho V^2 +p =\rho_0v_0^2+p_0.
\label{eq:app2}
\end{equation}
After crossing a shock front, the gas looses the thermal energy via radiation following the energy equation,
\begin{equation}
\frac{de}{dt}=-p\frac{d}{dt}\left(\frac{1}{\rho}\right)-\frac{{\Lambda}_{\rm net}}{\rho},
\end{equation}
where $e$ is the specific internal energy, $d/dt$ the Lagrangian time derivative,  
and $\Lambda_{\rm net}$ the net cooling rate per unit volume (in units of erg s$^{-1}$ cm$^{-3}$).
Assuming a strong shock (i.e. $\rho_0v_0^2\gg p_0$), 
the initial temperature and density for the post-shock flow are given by equations (\ref{eq:post_T}) and (\ref{eq:post_n})
using the velocity and density of the pre-shock flow.

As long as the cooling time of the shocked gas is shorter than the free-fall time
($t_{\rm ff}=\sqrt{3\pi/32G\rho}$),
which is the growth time-scale for gravitational instability \citep[e.g.][]{1985MNRAS.214..379L}, 
we follow the thermal evolution solving the equations (\ref{eq:app1}) and (\ref{eq:app2}).
When the cooling becomes inefficient and $t_{\rm cool}$ exceeds $t_{\rm ff}$, 
the contraction of the layer halts and dense clouds begin to develop inside the post-shock region. 
Once the mass exceeds the Jeans limit, the cloud collapses owing to its self-gravity in a runaway fashion 
following the equation
\begin{equation}
\frac{d\rho}{dt}=\frac{\rho}{t_{\rm ff}}.
\end{equation}
We consistently solve the chemical reaction networks among primordial species 
(H, H$_2$, e$^-$, H$^+$, H$_2^+$, H$^-$, He, He$^+$ and He$^{++}$).
We consider radiative cooling by free-free emission, atomic lines (H, He, He$^+$) and 
H$_2$ lines as well as chemical cooling/heating.

\begin{figure}
\begin{center}
\includegraphics[height=60mm,width=80mm]{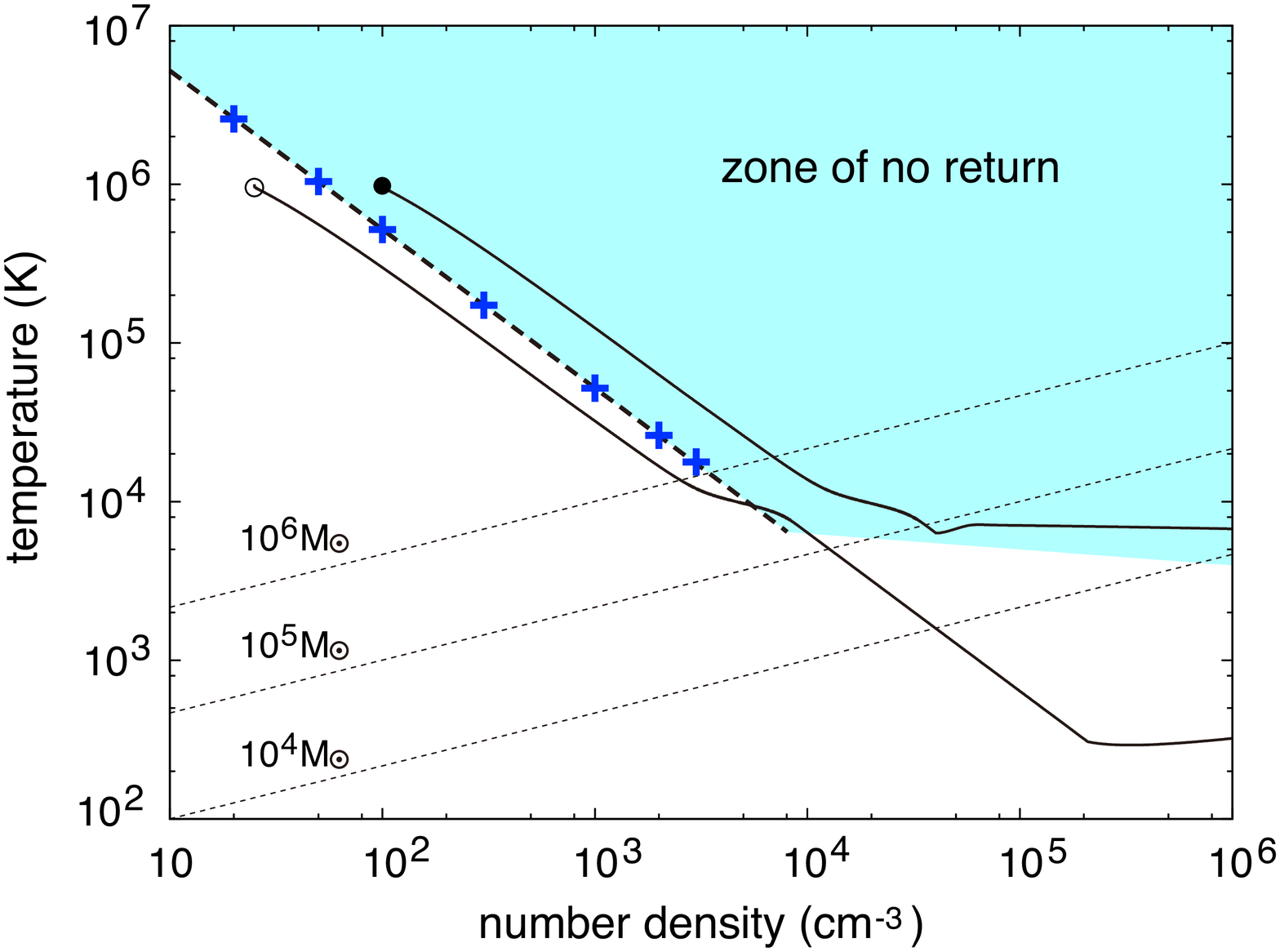}
\caption{Density and temperature conditions required for shocked gas to form supermassive clouds.
The solid curve with an open circle shows the thermal evolution of shocked gas with a lower initial density, 
and the solid curve with a filled circle shows the evolution for a higher initial density. 
The zone of no return is shown by the shaded region. 
Its boundary is given by cross symbols (numerical results) and the dashed line (fit given by equation \ref{eq:zone}).
The diagonal dotted lines indicate constant Jeans masses.}
\label{nt}
\end{center}
\end{figure}

Fig.~\ref{nt} shows the thermal evolutionary tracks of the post-shock gas. 
The solid curve which starts from the open (filled) circle corresponds to the case where the initial conditions of the post-shocks gas are $n=25~(100)~\cc$ and $T=10^6~{\rm K}$.
The corresponding density and velocity of the pre-shock flow are
$n_0=6.3~(25)~\cc$ and $v_0=270~{\rm km~s}^{-1}$, respectively.
Initially, hydrogen is fully ionized and helium is neutral.
For both cases, the gas cools down to $\sim 10^4$ K by free-free emission and 
atomic line emission (He$^+$, He, and H).
In the case of the lower initial density (open circle), the gas temperature decreases further to $\sim 300$ K by H$_2$-line cooling.
On the other hand, in the case of the higher initial density (filled circle), H$_2$ formation is suppressed by the collisional dissociation (${\rm H}+{\rm H}_2\rightarrow 3{\rm H}$).
The cooling becomes irrelevant when $t_{\rm cool}\ga t_{\rm ff}$. 
Then, the gas cloud collapses by the self gravity once the corresponding Jeans mass is assembled. 
In the case of the lower initial density, $t_{\rm cool}\ga t_{\rm ff}$ occurs at $n\simeq 2 \times 10^5~\cc$ and $T\sim 100$ K, 
and the corresponding Jeans mass is a few $\sim 100~\msun$. 
On the other hand, in the case of larger initial density, $t_{\rm cool}\ga t_{\rm ff}$ occurs at $n\simeq 4\times 10^4~\cc$ and $T\sim 8000$ K, 
and the corresponding Jeans mass is $\ga M_{\rm J}\simeq 10^5~\msun$. 
Such a massive cloud collapses almost isothermally, mediated by H atomic cooling, 
and forms a proto-SMS at the center without a major episode of fragmentation and subsequent H$_2$ formation
\citep{2014MNRAS.445L.109I,2015MNRAS.446.2380B}.

The shaded region in Fig~\ref{nt} represents the ``zone of no return".
If the gas jumps into this region by a strong shock, 
H$_2$ formation is suppressed by the collisional dissociation and massive clouds with $\ga 10^5~\msun$ are form.
The boundary is identified numerically (cross symbols),  
and the dashed line is the fitting, $T\ga 5.2 \times 10^5~{\rm K}~(n/10^2~\cc)^{-1}$ for $n\la 10^4~\cc$ (equation \ref{eq:zone}).

Finally, we note the dependence of the density range of the zone of no
return on the reaction rate coefficient of H$_2$ collisional
dissociation. We adopt the rate given by \cite{1996ApJ...461..265M}. 
The value of this coefficient is $\sim 30$ times higher than that
used in some previous studies at $n=10^3-10^4~\cc$
\citep[e.g.][]{1987ApJ...318...32S,2003ApJ...586....1M}. 
This results in an order of magnitude decrease in the density required to enter the zone of no
return and explains why these previous studies did not find gas
entering it.

\small{
\bibliography{ref.bib}

\begin{thebibliography}{83}
\expandafter\ifx\csname natexlab\endcsname\relax\def\natexlab#1{#1}\fi

\bibitem[{{Ade} {et~al}\mbox{.}(2014){Ade}, {Aghanim}, {Armitage-Caplan},
  {Arnaud}, {Ashdown}, {Atrio-Barandela}, {Aumont}, {Baccigalupi}, {Banday}, \&
  et~al.}]{2014A&A...571A..16P}
{Ade} P.~A.~R. {et~al.}, 2014, \aap, 571, A16

\bibitem[{{Ahn} {et~al}\mbox{.}(2009){Ahn}, {Shapiro}, {Iliev}, {Mellema}, \&
  {Pen}}]{2009ApJ...695.1430A}
{Ahn} K., {Shapiro} P.~R., {Iliev} I.~T., {Mellema} G., {Pen} U.-L., 2009,
  \apj, 695, 1430

\bibitem[{{Barkana} \& {Loeb}(2001)}]{2001PhR...349..125B}
{Barkana} R., {Loeb} A., 2001, \physrep, 349, 125

\bibitem[{{Becerra} {et~al}\mbox{.}(2015){Becerra}, {Greif}, {Springel}, \&
  {Hernquist}}]{2015MNRAS.446.2380B}
{Becerra} F., {Greif} T.~H., {Springel} V., {Hernquist} L.~E., 2015, \mnras,
  446, 2380

\bibitem[{{Begelman}, {Volonteri} \& {Rees}(2006){Begelman}, {Volonteri}, \&
  {Rees}}]{2006MNRAS.370..289B}
{Begelman} M.~C., {Volonteri} M., {Rees} M.~J., 2006, \mnras, 370, 289

\bibitem[{{Behroozi}, {Wechsler} \& {Wu}(2013){Behroozi}, {Wechsler}, \&
  {Wu}}]{2013ApJ...762..109B}
{Behroozi} P.~S., {Wechsler} R.~H., {Wu} H.-Y., 2013, \apj, 762, 109

\bibitem[{{Bromm} \& {Loeb}(2003)}]{2003ApJ...596...34B}
{Bromm} V., {Loeb} A., 2003, \apj, 596, 34

\bibitem[{{Chandrasekhar}(1964)}]{1964ApJ...140..417C}
{Chandrasekhar} S., 1964, \apj, 140, 417

\bibitem[{{Coles} \& {Jones}(1991)}]{1991MNRAS.248....1C}
{Coles} P., {Jones} B., 1991, \mnras, 248, 1

\bibitem[{{Condon}, {Helou} \& {Jarrett}(2002){Condon}, {Helou}, \&
  {Jarrett}}]{2002AJ....123.1881C}
{Condon} J.~J., {Helou} G., {Jarrett} T.~H., 2002, \aj, 123, 1881

\bibitem[{{Condon} {et~al}\mbox{.}(1993){Condon}, {Helou}, {Sanders}, \&
  {Soifer}}]{1993AJ....105.1730C}
{Condon} J.~J., {Helou} G., {Sanders} D.~B., {Soifer} B.~T., 1993, \aj, 105,
  1730

\bibitem[{{Dijkstra}, {Ferrara} \& {Mesinger}(2014){Dijkstra}, {Ferrara}, \&
  {Mesinger}}]{Dijkstra+14}
{Dijkstra} M., {Ferrara} A., {Mesinger} A., 2014, \mnras, 442, 2036

\bibitem[{{Fan}(2006)}]{2006NewAR..50..665F}
{Fan} X., 2006, \nar, 50, 665

\bibitem[{{Fernandez} {et~al}\mbox{.}(2014){Fernandez}, {Bryan}, {Haiman}, \&
  {Li}}]{2014MNRAS.439.3798F}
{Fernandez} R., {Bryan} G.~L., {Haiman} Z., {Li} M., 2014, \mnras, 439, 3798

\bibitem[{{Ferrara}, {Haardt} \& {Salvaterra}(2013){Ferrara}, {Haardt}, \&
  {Salvaterra}}]{2013MNRAS.434.2600F}
{Ferrara} A., {Haardt} F., {Salvaterra} R., 2013, \mnras, 434, 2600

\bibitem[{{Field}(1965)}]{1965ApJ...142..531F}
{Field} G.~B., 1965, \apj, 142, 531

\bibitem[{{Glover} \& {Jappsen}(2007)}]{2007ApJ...666....1G}
{Glover} S.~C.~O., {Jappsen} A.-K., 2007, \apj, 666, 1

\bibitem[{{Greif} {et~al}\mbox{.}(2010){Greif}, {Glover}, {Bromm}, \&
  {Klessen}}]{2010ApJ...716..510G}
{Greif} T.~H., {Glover} S.~C.~O., {Bromm} V., {Klessen} R.~S., 2010, \apj, 716,
  510

\bibitem[{{Haiman}, {Abel} \& {Rees}(2000){Haiman}, {Abel}, \&
  {Rees}}]{2000ApJ...534...11H}
{Haiman} Z., {Abel} T., {Rees} M.~J., 2000, \apj, 534, 11

\bibitem[{{Hamana} {et~al}\mbox{.}(2003){Hamana}, {Kayo}, {Yoshida}, {Suto}, \&
  {Jing}}]{2003MNRAS.343.1312H}
{Hamana} T., {Kayo} I., {Yoshida} N., {Suto} Y., {Jing} Y.~P., 2003, \mnras,
  343, 1312

\bibitem[{{Heger} \& {Woosley}(2002)}]{2002ApJ...567..532H}
{Heger} A., {Woosley} S.~E., 2002, \apj, 567, 532

\bibitem[{{Herrmann} {et~al}\mbox{.}(2007){Herrmann}, {Hinder}, {Shoemaker},
  {Laguna}, \& {Matzner}}]{2007ApJ...661..430H}
{Herrmann} F., {Hinder} I., {Shoemaker} D., {Laguna} P., {Matzner} R.~A., 2007,
  \apj, 661, 430

\bibitem[{{Hirano} {et~al}\mbox{.}(2014){Hirano}, {Hosokawa}, {Yoshida},
  {Umeda}, {Omukai}, {Chiaki}, \& {Yorke}}]{2014ApJ...781...60H}
{Hirano} S., {Hosokawa} T., {Yoshida} N., {Umeda} H., {Omukai} K., {Chiaki} G.,
  {Yorke} H.~W., 2014, \apj, 781, 60

\bibitem[{{Hosokawa}, {Omukai} \& {Yorke}(2012){Hosokawa}, {Omukai}, \&
  {Yorke}}]{2012ApJ...756...93H}
{Hosokawa} T., {Omukai} K., {Yorke} H.~W., 2012, \apj, 756, 93

\bibitem[{{Hosokawa} {et~al}\mbox{.}(2011){Hosokawa}, {Omukai}, {Yoshida}, \&
  {Yorke}}]{2011Sci...334.1250H}
{Hosokawa} T., {Omukai} K., {Yoshida} N., {Yorke} H.~W., 2011, Science, 334,
  1250

\bibitem[{{Hosokawa} {et~al}\mbox{.}(2013){Hosokawa}, {Yorke}, {Inayoshi},
  {Omukai}, \& {Yoshida}}]{2013ApJ...778..178H}
{Hosokawa} T., {Yorke} H.~W., {Inayoshi} K., {Omukai} K., {Yoshida} N., 2013,
  \apj, 778, 178

\bibitem[{{Hosokawa} {et~al}\mbox{.}(2012){Hosokawa}, {Yoshida}, {Omukai}, \&
  {Yorke}}]{2012ApJ...760L..37H}
{Hosokawa} T., {Yoshida} N., {Omukai} K., {Yorke} H.~W., 2012, \apjl, 760, L37

\bibitem[{{Inayoshi} \& {Omukai}(2011)}]{IO11}
{Inayoshi} K., {Omukai} K., 2011, \mnras, 416, 2748

\bibitem[{{Inayoshi} \& {Omukai}(2012)}]{IO12}
---, 2012, \mnras, 422, 2539

\bibitem[{{Inayoshi}, {Omukai} \& {Tasker}(2014){Inayoshi}, {Omukai}, \&
  {Tasker}}]{2014MNRAS.445L.109I}
{Inayoshi} K., {Omukai} K., {Tasker} E., 2014, \mnras, 445, L109

\bibitem[{{Inayoshi} \& {Tanaka}(2014)}]{2014arXiv1411.2590I}
{Inayoshi} K., {Tanaka} T.~L., 2014, arXiv: 1411.2590

\bibitem[{{Johnson} \& {Bromm}(2007)}]{2007MNRAS.374.1557J}
{Johnson} J.~L., {Bromm} V., 2007, \mnras, 374, 1557

\bibitem[{{Johnson}, {Dalla} \& {Khochfar}(2013){Johnson}, {Dalla}, \&
  {Khochfar}}]{2013MNRAS.428.1857J}
{Johnson} J.~L., {Dalla} V.~C., {Khochfar} S., 2013, \mnras, 428, 1857

\bibitem[{{Johnson} {et~al}\mbox{.}(2011){Johnson}, {Khochfar}, {Greif}, \&
  {Durier}}]{Johnson+11}
{Johnson} J.~L., {Khochfar} S., {Greif} T.~H., {Durier} F., 2011, \mnras, 410,
  919

\bibitem[{{Koppitz} {et~al}\mbox{.}(2007){Koppitz}, {Pollney}, {Reisswig},
  {Rezzolla}, {Thornburg}, {Diener}, \& {Schnetter}}]{2007PhRvL..99d1102K}
{Koppitz} M., {Pollney} D., {Reisswig} C., {Rezzolla} L., {Thornburg} J.,
  {Diener} P., {Schnetter} E., 2007, Physical Review Letters, 99, 041102

\bibitem[{{Koushiappas}, {Bullock} \& {Dekel}(2004){Koushiappas}, {Bullock}, \&
  {Dekel}}]{2004MNRAS.354..292K}
{Koushiappas} S.~M., {Bullock} J.~S., {Dekel} A., 2004, \mnras, 354, 292

\bibitem[{{Larson}(1985)}]{1985MNRAS.214..379L}
{Larson} R.~B., 1985, \mnras, 214, 379

\bibitem[{{Larson} \& {Tinsley}(1978)}]{1978ApJ...219...46L}
{Larson} R.~B., {Tinsley} B.~M., 1978, \apj, 219, 46

\bibitem[{{Latif} {et~al}\mbox{.}(2014){Latif}, {Bovino}, {Van Borm}, {Grassi},
  {Schleicher}, \& {Spaans}}]{2014MNRAS.443.1979L}
{Latif} M.~A., {Bovino} S., {Van Borm} C., {Grassi} T., {Schleicher} D.~R.~G.,
  {Spaans} M., 2014, \mnras, 443, 1979

\bibitem[{{Latif} {et~al}\mbox{.}(2013){Latif}, {Schleicher}, {Schmidt}, \&
  {Niemeyer}}]{2013MNRAS.433.1607L}
{Latif} M.~A., {Schleicher} D.~R.~G., {Schmidt} W., {Niemeyer} J., 2013,
  \mnras, 433, 1607

\bibitem[{{Loeb} \& {Rasio}(1994)}]{1994ApJ...432...52L}
{Loeb} A., {Rasio} F.~A., 1994, \apj, 432, 52

\bibitem[{{Machacek}, {Bryan} \& {Abel}(2001){Machacek}, {Bryan}, \&
  {Abel}}]{2001ApJ...548..509M}
{Machacek} M.~E., {Bryan} G.~L., {Abel} T., 2001, \apj, 548, 509

\bibitem[{{Mackey}, {Bromm} \& {Hernquist}(2003){Mackey}, {Bromm}, \&
  {Hernquist}}]{2003ApJ...586....1M}
{Mackey} J., {Bromm} V., {Hernquist} L., 2003, \apj, 586, 1

\bibitem[{{Maio} {et~al}\mbox{.}(2010){Maio}, {Ciardi}, {Dolag}, {Tornatore},
  \& {Khochfar}}]{2010MNRAS.407.1003M}
{Maio} U., {Ciardi} B., {Dolag} K., {Tornatore} L., {Khochfar} S., 2010,
  \mnras, 407, 1003

\bibitem[{{Markevitch} {et~al}\mbox{.}(2002){Markevitch}, {Gonzalez}, {David},
  {Vikhlinin}, {Murray}, {Forman}, {Jones}, \& {Tucker}}]{2002ApJ...567L..27M}
{Markevitch} M., {Gonzalez} A.~H., {David} L., {Vikhlinin} A., {Murray} S.,
  {Forman} W., {Jones} C., {Tucker} W., 2002, \apjl, 567, L27

\bibitem[{{Martin}, {Schwarz} \& {Mandy}(1996){Martin}, {Schwarz}, \&
  {Mandy}}]{1996ApJ...461..265M}
{Martin} P.~G., {Schwarz} D.~H., {Mandy} M.~E., 1996, \apj, 461, 265

\bibitem[{{Mayer} {et~al}\mbox{.}(2014){Mayer}, {Fiacconi}, {Bonoli}, {Quinn},
  {Roskar}, {Shen}, \& {Wadsley}}]{Mayer_et_al_2014}
{Mayer} L., {Fiacconi} D., {Bonoli} S., {Quinn} T., {Roskar} R., {Shen} S.,
  {Wadsley} J., 2014, ArXiv e-prints

\bibitem[{{Mayer} {et~al}\mbox{.}(2010){Mayer}, {Kazantzidis}, {Escala}, \&
  {Callegari}}]{2010Natur.466.1082M}
{Mayer} L., {Kazantzidis} S., {Escala} A., {Callegari} S., 2010, \nat, 466,
  1082

\bibitem[{{Milosavljevi{\'c}}, {Couch} \& {Bromm}(2009){Milosavljevi{\'c}},
  {Couch}, \& {Bromm}}]{2009ApJ...696L.146M}
{Milosavljevi{\'c}} M., {Couch} S.~M., {Bromm} V., 2009, \apjl, 696, L146

\bibitem[{{Mortlock} {et~al}\mbox{.}(2011){Mortlock}, {Warren}, {Venemans},
  {Patel}, {Hewett}, {McMahon}, {Simpson}, {Theuns}, {Gonz{\'a}les-Solares},
  {Adamson}, {Dye}, {Hambly}, {Hirst}, {Irwin}, {Kuiper}, {Lawrence}, \&
  {R{\"o}ttgering}}]{2011Natur.474..616M}
{Mortlock} D.~J. {et~al.}, 2011, \nat, 474, 616

\bibitem[{{Nomoto} {et~al}\mbox{.}(2006){Nomoto}, {Tominaga}, {Umeda},
  {Kobayashi}, \& {Maeda}}]{2006NuPhA.777..424N}
{Nomoto} K., {Tominaga} N., {Umeda} H., {Kobayashi} C., {Maeda} K., 2006,
  Nuclear Physics A, 777, 424

\bibitem[{{Oh} \& {Haiman}(2002)}]{2002ApJ...569..558O}
{Oh} S.~P., {Haiman} Z., 2002, \apj, 569, 558

\bibitem[{{Omukai}(2001)}]{O01}
{Omukai} K., 2001, \apj, 546, 635

\bibitem[{{O'Shea} \& {Norman}(2008)}]{2008ApJ...673...14O}
{O'Shea} B.~W., {Norman} M.~L., 2008, \apj, 673, 14

\bibitem[{{Pallottini} {et~al}\mbox{.}(2014){Pallottini}, {Ferrara},
  {Gallerani}, {Salvadori}, \& {D'Odorico}}]{2014MNRAS.440.2498P}
{Pallottini} A., {Ferrara} A., {Gallerani} S., {Salvadori} S., {D'Odorico} V.,
  2014, \mnras, 440, 2498

\bibitem[{{Park} \& {Ricotti}(2011)}]{2011ApJ...739....2P}
{Park} K., {Ricotti} M., 2011, \apj, 739, 2

\bibitem[{{Park} \& {Ricotti}(2012)}]{2012ApJ...747....9P}
---, 2012, \apj, 747, 9

\bibitem[{{Regan} \& {Haehnelt}(2009{\natexlab{a}})}]{2009MNRAS.396..343R}
{Regan} J.~A., {Haehnelt} M.~G., 2009{\natexlab{a}}, \mnras, 396, 343

\bibitem[{{Regan} \& {Haehnelt}(2009{\natexlab{b}})}]{2009MNRAS.393..858R}
---, 2009{\natexlab{b}}, \mnras, 393, 858

\bibitem[{{Saitoh} {et~al}\mbox{.}(2009){Saitoh}, {Daisaka}, {Kokubo},
  {Makino}, {Okamoto}, {Tomisaka}, {Wada}, \& {Yoshida}}]{Saitoh_et_al_2009}
{Saitoh} T.~R., {Daisaka} H., {Kokubo} E., {Makino} J., {Okamoto} T.,
  {Tomisaka} K., {Wada} K., {Yoshida} N., 2009, \pasj, 61, 481

\bibitem[{{Schleicher} {et~al}\mbox{.}(2013){Schleicher}, {Palla}, {Ferrara},
  {Galli}, \& {Latif}}]{2013A&A...558A..59S}
{Schleicher} D.~R.~G., {Palla} F., {Ferrara} A., {Galli} D., {Latif} M., 2013,
  \aap, 558, A59

\bibitem[{{Shang}, {Bryan} \& {Haiman}(2010){Shang}, {Bryan}, \&
  {Haiman}}]{2010MNRAS.402.1249S}
{Shang} C., {Bryan} G.~L., {Haiman} Z., 2010, \mnras, 402, 1249

\bibitem[{{Shapiro} \& {Kang}(1987)}]{1987ApJ...318...32S}
{Shapiro} P.~R., {Kang} H., 1987, \apj, 318, 32

\bibitem[{{Sheth} \& {Diaferio}(2001)}]{2001MNRAS.322..901S}
{Sheth} R.~K., {Diaferio} A., 2001, \mnras, 322, 901

\bibitem[{{Sheth} \& {Tormen}(1999)}]{1999MNRAS.308..119S}
{Sheth} R.~K., {Tormen} G., 1999, \mnras, 308, 119

\bibitem[{{Shibata} \& {Shapiro}(2002)}]{2002ApJ...572L..39S}
{Shibata} M., {Shapiro} S.~L., 2002, \apjl, 572, L39

\bibitem[{{Springel}(2005)}]{2005MNRAS.364.1105S}
{Springel} V., 2005, \mnras, 364, 1105

\bibitem[{{Stacy}, {Greif} \& {Bromm}(2012){Stacy}, {Greif}, \&
  {Bromm}}]{2012MNRAS.422..290S}
{Stacy} A., {Greif} T.~H., {Bromm} V., 2012, \mnras, 422, 290

\bibitem[{{Sugimura}, {Omukai} \& {Inoue}(2014){Sugimura}, {Omukai}, \&
  {Inoue}}]{2014MNRAS.445..544S}
{Sugimura} K., {Omukai} K., {Inoue} A.~K., 2014, \mnras, 445, 544

\bibitem[{{Susa}, {Hasegawa} \& {Tominaga}(2014){Susa}, {Hasegawa}, \&
  {Tominaga}}]{2014ApJ...792...32S}
{Susa} H., {Hasegawa} K., {Tominaga} N., 2014, \apj, 792, 32

\bibitem[{{Sutherland} \& {Dopita}(1993)}]{1993ApJS...88..253S}
{Sutherland} R.~S., {Dopita} M.~A., 1993, \apjs, 88, 253

\bibitem[{{Tanaka} \& {Haiman}(2009)}]{TH09}
{Tanaka} T., {Haiman} Z., 2009, \apj, 696, 1798

\bibitem[{{Tanaka}, {Perna} \& {Haiman}(2012){Tanaka}, {Perna}, \&
  {Haiman}}]{2012MNRAS.425.2974T}
{Tanaka} T., {Perna} R., {Haiman} Z., 2012, \mnras, 425, 2974

\bibitem[{{Tornatore}, {Ferrara} \& {Schneider}(2007){Tornatore}, {Ferrara}, \&
  {Schneider}}]{2007MNRAS.382..945T}
{Tornatore} L., {Ferrara} A., {Schneider} R., 2007, \mnras, 382, 945

\bibitem[{{Tucker} {et~al}\mbox{.}(1998){Tucker}, {Blanco}, {Rappoport},
  {David}, {Fabricant}, {Falco}, {Forman}, {Dressler}, \&
  {Ramella}}]{1998ApJ...496L...5T}
{Tucker} W. {et~al.}, 1998, \apjl, 496, L5

\bibitem[{{Visbal}, {Haiman} \& {Bryan}(2014{\natexlab{a}}){Visbal}, {Haiman},
  \& {Bryan}}]{Visbal+14}
{Visbal} E., {Haiman} Z., {Bryan} G.~L., 2014{\natexlab{a}}, \mnras, 442, L100

\bibitem[{{Visbal}, {Haiman} \& {Bryan}(2014{\natexlab{b}}){Visbal}, {Haiman},
  \& {Bryan}}]{2014MNRAS.445.1056V}
---, 2014{\natexlab{b}}, \mnras, 445, 1056

\bibitem[{{Visbal} {et~al}\mbox{.}(2014){Visbal}, {Haiman}, {Terrazas},
  {Bryan}, \& {Barkana}}]{2014MNRAS.445..107V}
{Visbal} E., {Haiman} Z., {Terrazas} B., {Bryan} G.~L., {Barkana} R., 2014,
  \mnras, 445, 107

\bibitem[{{Willott} {et~al}\mbox{.}(2010){Willott}, {Delorme}, {Reyl{\'e}},
  {Albert}, {Bergeron}, {Crampton}, {Delfosse}, {Forveille}, {Hutchings},
  {McLure}, {Omont}, \& {Schade}}]{2010AJ....139..906W}
{Willott} C.~J. {et~al.}, 2010, \aj, 139, 906

\bibitem[{{Wise} \& {Abel}(2007)}]{2007ApJ...671.1559W}
{Wise} J.~H., {Abel} T., 2007, \apj, 671, 1559

\bibitem[{{Wise} {et~al}\mbox{.}(2012){Wise}, {Turk}, {Norman}, \&
  {Abel}}]{2012ApJ...745...50W}
{Wise} J.~H., {Turk} M.~J., {Norman} M.~L., {Abel} T., 2012, \apj, 745, 50

\bibitem[{{Wu} {et~al}\mbox{.}(2015){Wu}, {Wang}, {Fan}, {Yi}, {Zuo}, {Bian},
  {Jiang}, {McGreer}, {Wang}, {Yang}, {Yang}, {Thompson}, \&
  {Beletsky}}]{2015Natur.518..512W}
{Wu} X.-B. {et~al.}, 2015, \nat, 518, 512

\bibitem[{{Zeldovich} \& {Novikov}(1971)}]{1971reas.book.....Z}
{Zeldovich} Y.~B., {Novikov} I.~D., 1971, {Relativistic astrophysics. Vol.1:
  Stars and relativity}

\end{thebibliography}
}
\end{document}